\documentclass[fleqn,a4]{article}
\usepackage{amsmath,amssymb,amscd}
\usepackage[pdftex]{graphicx}
\makeatletter

\selectfont

\renewcommand{\section}{\@startsection{section}{1}{\z@}%
{2ex}{1ex}{\reset@font\large\bfseries}}%
\renewcommand{\thesection}{\@arabic\c@section}
\def\@listi{\topsep=.3\baselineskip \parsep=.2ex \partopsep=0ex%
\itemsep=0ex \leftmargin=4ex \rightmargin=2ex}
\let\@listI\@listi
\@listi\def\@listii{\parsep=.2ex \partopsep=0pt \itemsep=0ex%
\leftmargin=4ex \rightmargin=0ex}
\let\@listiii\@listii
\let\@listiv\@listii
\let\@listv\@listii
\let\@listvi\@listii
\long\def\@makecaption#1#2{\footnotesize\sbox\@tempboxa{#1. #2}
\ifdim\wd\@tempboxa >\hsize #1. #2\par
\else \global\@minipagefalse
\hb@xt@\hsize{\hfil\box\@tempboxa\hfil}
\fi}
\makeatother

\newtheorem{theorem}{Theorem}

\usepackage{graphicx}
\usepackage{amsmath}
\usepackage{url}

\begin{document}
\title{On a parametrized difference equation connecting chaotic and integrable mappings}
\author{Tomoko Nagai$^{1}$\thanks{kt13643@ns.kogakuin.ac.jp},
Atsushi Nagai$^2$, 
Hiroko Yamaki$^2$, and Kana Yanuma$^2$ \\
$^{1}$ Academic Support Center, Kogakuin University, \\2665-1 Nakano-cho, 
Hachioji, Tokyo 192-0015, Japan \\ 
$^2$Department of Computer Sciences, Tsuda University, \\
2-1-1 Tsuda-cho, Kodaira, Tokyo 187-8577, Japan}
\date{\empty}


\maketitle
\begin{abstract}
We present a new difference equation with two parameters $c \in [0,1]$ and $A \in [1,4]$. This equation
is equivalent to the logistic mapping if $c=1$ and the Morishita mapping if $c=0$, which are the well-known 
chaotic and integrable mappings, respectively. We first consider the case $A=4$ and investigate the time evolution by 
changing the parameter $c\in [0,1]$. We next change both two parameters $A \in [3,4]$ and $c \in [0,1]$ and 
present the corresponding 3D bifurcation diagram.
\end{abstract}
\section{Introduction}
The logistic equation, 
\begin{align}
\frac{du}{dt} = a u(1-u)\qquad u(0)=u_0, 
\end{align}
where $a$ is a positive constant and $u=u(t)$ is an unknown function, 
is a model equation describing population dynamics and possesses a solution
\begin{align}
u(t)=\frac{u_0e^{at}}{u_0e^{at}+1-u_0}
\end{align}
Concerning its discrete version, the following two difference equations are well-known: 
\begin{align}
& u_{n+1}= A u_n(1-u_n),\label{eq:logistic} \\
& u_{n+1} = A u_n(1-u_{n+1}) \Leftrightarrow u_{n+1}=\frac{A u_n}{1+A u_n},\label{eq:morisita}
\end{align}
where $A \in (1,4]$ is a given constant.

Eq.~\eqref{eq:logistic} is the well-known logistic mapping and 
exhibits a chaotic behavior if $A$ exceeds the value $A=3.5699456\cdots$.(See \cite{May}, for example.)

On the other hand, eq.~\eqref{eq:morisita}, which is called the Morishita mapping, is an integrable 
mapping \cite{M, HT}. In other words, 
eq.~\eqref{eq:morisita} is linearized by taking its reciprocal and putting $v_n=1/u_n$ as 
\begin{align*}
v_{n+1}=\frac1{A}v_n+1. 
\end{align*}
This is solved as 
\begin{align*}
v_n=\frac{A}{A-1}+\frac1{A^n}\left(\frac1{u_0}-\frac{A}{A-1}\right)
\end{align*}
and therefore $u_n$ is given by
\begin{align}
u_n=\frac{A^n(A-1)u_0}{A(A^n-1)u_0+A-1},
\end{align}
which converges to $\dfrac{A-1}{A}$ as $n$ tends to $\infty$.

The purpose of this paper is to present a new difference equation connecting the above two 
different mappings. We also investigate the time evolution of this new equation and 
calculate the Lyapunov exponent. The bifurcation diagram is also presented.

\section{A new difference equation connecting the logistic and Morishita mappings}
In this section, we put $A=4$ and consider a new mapping including a variable parameter 
$c \in [0,1]$, 
\begin{align}
& u_{n+1}=4u_n(1-cu_n-(1-c)u_{n+1}) \nonumber \\
\Leftrightarrow ~& u_{n+1}=\frac{4u_n(1-cu_n)}{1+4(1-c)u_n}=f(c,u_n),
\label{eq:a}
\end{align}
where $f(c,x)=\dfrac{4x(1-cx)}{1+4(1-c)x}$. 
It is easy to observe that the mapping \eqref{eq:a} is equivalent to the Morishita mapping if $c=0$ and 
to the logistic mapping if $c=1$. 

Concerning the properties of $f(c,x)$, we have the following theorem.
\begin{theorem}
The function $f(c,x)\quad (0\leq x\leq 1,~0\leq c \leq 1)$ satisfies the following properties.
\begin{align}
& f(c,\frac34)=\frac34 \label{eq:th1} \\
& 0\leq f(c,x) \leq 1 \label{eq:th2}
\end{align}
\end{theorem}
The relation \eqref{eq:th1} means that $u_n=\frac34$ is an equilibrium point of the mapping~\eqref{eq:a}
and \eqref{eq:th2} means that if $u_0 \in [0,1]$ we have $u_n \in [0,1]$ for any $n =1,2,\cdots$.\\
{\bf Proof~:}~
The relation~\eqref{eq:th1} is easily shown through direct calculation. 
In order to prove \eqref{eq:th2}, we take a derivative of $f(c,x)$ 
with respect to $c$, which is given by
\begin{align*}
\frac{\partial}{\partial c} f(c,x)=\frac{4x^2(3-4x)}{(1+4x-4cx)^2}.
\end{align*}
Hence we have 
\begin{align*}
f(0,x) \leq f(c,x) \leq f(1,x) &\qquad (0 \leq x \leq \frac34), \\
f(1,x) \leq f(c,x) \leq f(0,x) &\qquad (\frac34 \leq x \leq 1).
\end{align*}
Together with the facts,
\begin{align*}
0\leq f(0,x)=\frac{4x}{1+4x}<1,~~~0\leq f(1,x)=4x(1-x) \leq 1,
\end{align*}
the relation \eqref{eq:th2} follows. \hfill $\blacksquare$

\begin{figure}[htbp]
\includegraphics[scale=0.5]{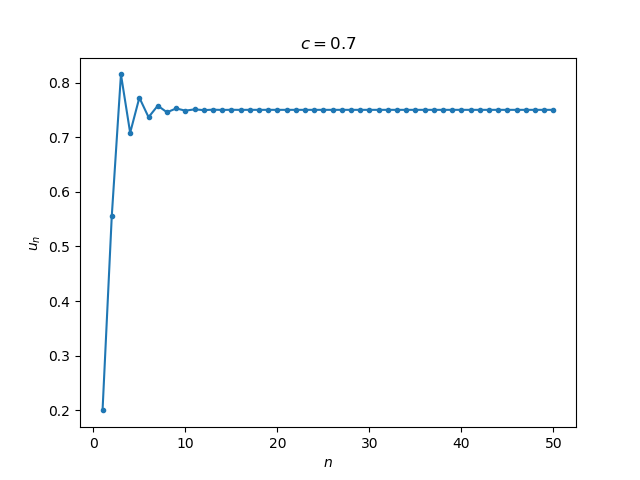}
\includegraphics[scale=0.5]{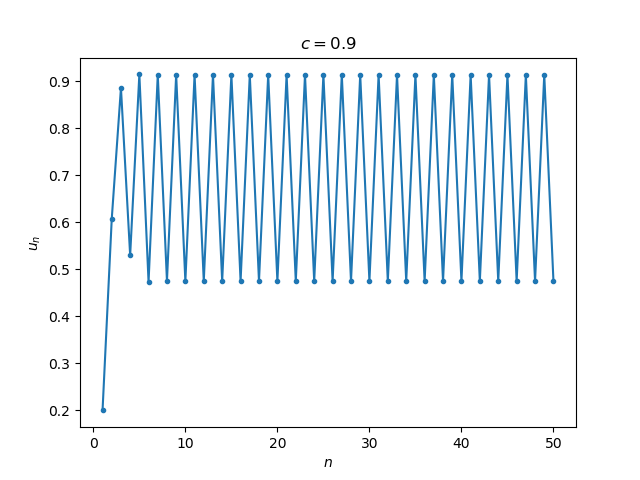}
\includegraphics[scale=0.5]{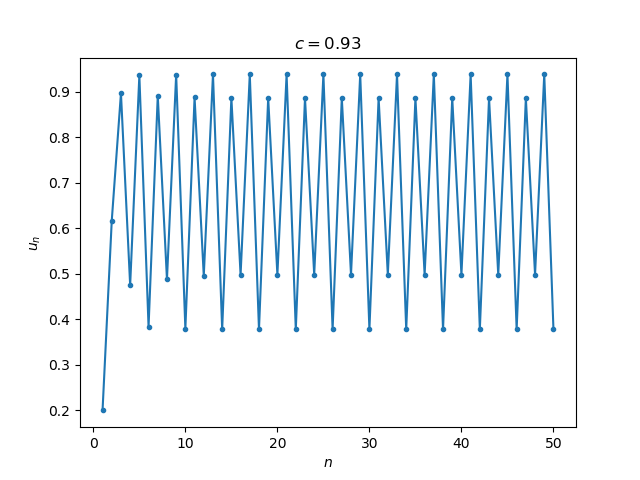}
\includegraphics[scale=0.5]{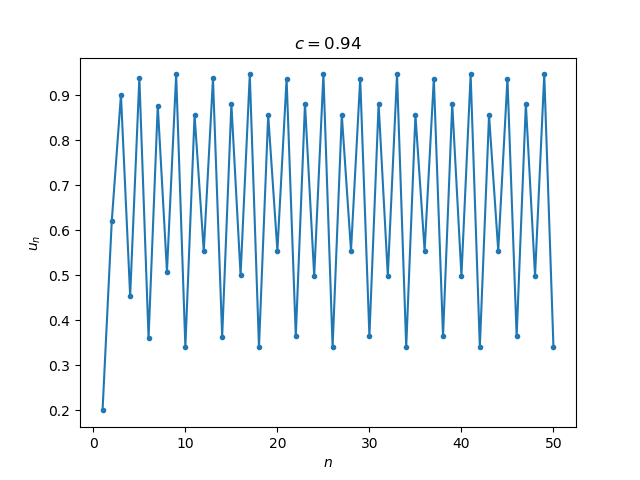}
\includegraphics[scale=0.5]{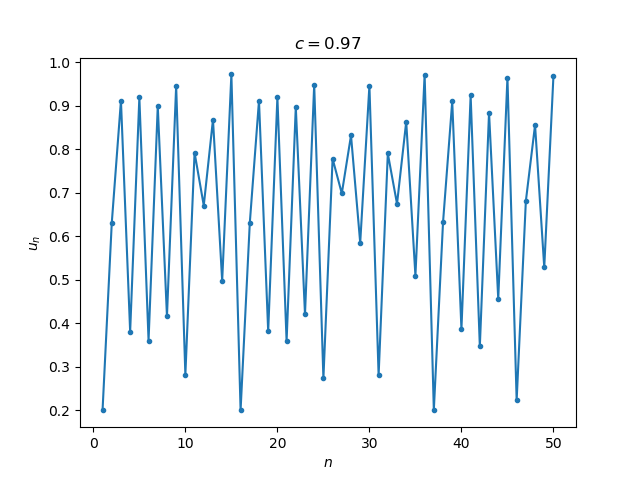}
\includegraphics[scale=0.5]{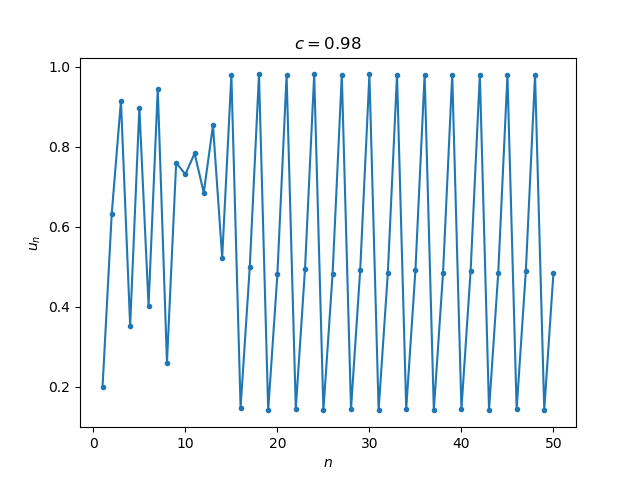}
\caption{Time evolutions of the mapping~\eqref{eq:a} for 
$c=0.7$,~$c=0.9$(period 2),~$c=0.93$(period 4),~
$c=0.94$(period 8),~$c=0.97$(chaotic),~$c=0.98$(period 3)}
\label{fig:time}
\end{figure}

By changing $c$ in an interval $[0,1]$, we calculate the time evolutions of $\{u_n\}$, which 
are given in Fig.~\ref{fig:time}. Numerical calculations are performed by Python 3.9.

We here investigate the Fig.~\ref{fig:time} in a detailed manner.
If $0\leq c < \frac56$, we can observe that $u_n$ converges to $\frac34$. This is confirmed as
follows. We  put $u_n=\frac34+\varepsilon$ in eq.~\eqref{eq:a} and have the Taylor expansion
of $u_{n+1}=f(c,u_n)$ around $\varepsilon=0$ as follows.
\begin{align*}
u_{n+1}&=\frac{(3+4\varepsilon)(1-c(\frac34+\varepsilon))}{1+(1-c)(3+4\varepsilon)}
=\frac34+\frac{-1+3c}{-4+3c}\varepsilon+\cdots.
\end{align*}
If $0\leq c < \frac56$ we have $\left| \frac{-1+3c}{-4+3c} \right|<1$  and therefore
the equilibrium point $u_n=\frac34$ is stable. 

If $c$ exceeds $\frac56$, we have 
$\left| \frac{-1+3c}{-4+3c} \right|>1$ and therefore $u_n=\frac34$ is unstable. 
It should be noted that $u_n=0$ is another equilibrium point, which is, however, 
unstable for $c \in [0,1]$.

Through straightforward calucations, if $\frac56 < c$, 
$u_n$ converges to a periodic orbit with period 2. That is, $u_n$ takes alternately two values
\begin{align*}
u_{2,\pm}=\frac{-5+10c\pm\sqrt5\,\sqrt{5-16c+12c^2}}{8(-c+2c^2)},
\end{align*} 
which are solutions of the equation $f(c,f(c,x))=x$ except $x=0,~\frac34$, 
as $n$ tends to $\infty$. In other words, we have 
\begin{align*}
f(c,u_{2,\pm})=u_{2,\mp}.
\end{align*}
The Taylor expansion of $f(c,f(c,u_{2,\pm}+\varepsilon))$ around
$\varepsilon=0$ is given by 
\begin{align*}
f(c,f(c,u_{2,\pm}+\varepsilon))=u_{2,\pm}+\frac{63c^2-84c+25}{3c^2-4c}\varepsilon+\cdots.
\end{align*}
Hence if $c$ exceeds the value $\frac{44+\sqrt{286}}{66}=0.922902\cdots$, 
which is a solution to $\left|\frac{63c^2-84c+25}{3c^2-4c}\right|=1$, the period of the sequence $\{u_n\}$ becomes 4. 

If we further increase the parameter $c$, the period of the sequence $\{u_n\}$ is 
doubled as $2^{3}, 2^{4}, 2^{5}, \cdots$ and finally a chaotic behavior appears at $c=0.942\cdots$. 
We next calculate the value $c_{n}$ where $u_{n}$ is $2^{n}$-periodic if $c_{n}<c<c_{n+1}$.
We further calculate 
\begin{align}\label{eq:Fn}
F_{n}=\frac{c_{n+1}-c_{n}}{c_{n+2}-c_{n+1}},
\end{align}
which is expected to converge to the Feigenbaum constant $4.669201\cdots$ as is shown in Table \ref{t1}.\\[2mm]
\begin{table}
\caption{$n$ dependence of $F_n$ in  Eq.~\eqref{eq:a}}
\label{t1}
\begin{center}
\begin{tabular}{lllllllll}
\hline
\multicolumn{1}{c}{$n$} & \multicolumn{1}{c}{1 } & \multicolumn{1}{c}{ 2 } & \multicolumn{1}{c}{3 } & \multicolumn{1}{c}{ 4 } & \multicolumn{1}{c}{5 } & \multicolumn{1}{c}{6 } & \multicolumn{1}{c}{7 } & \multicolumn{1}{c}{8 } \\
\hline
$c_{n}$ & 0.833333 & 0.922902 & 0.93803 & 0.941167 & 0.941833 & 0.9419755 & 0.94200603 & 0.94201257\\
$F_{n}$ & 5.920743 & 4.822442& 4.710210 & 4.657343 & 4.761904 & 4.668196 & & \\
\hline
\end{tabular}
\end{center}
\end{table}

It is interesting to note that the period 3 appears around $c_{3}^{(1)}<c<c_{3}^{(2)}$, 
which reminds us of Li and 
York's famous result~\cite{Li}.
The approximated critical values are given by $c_{3}^{(1)}=0.978471\cdots$ and 
$c_{3}^{(2)}=0.980113\cdots$.

\begin{figure}
\begin{center}
\includegraphics[scale=0.5]{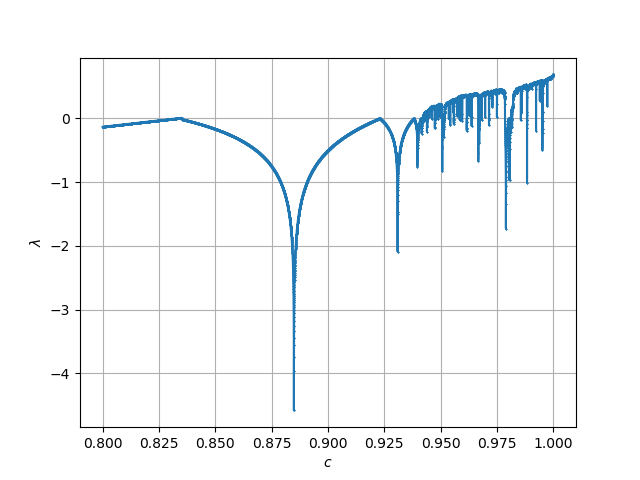}
\end{center}
\caption{Lyapunov exponent}
\label{fig:lyapunov}
\end{figure}

In Fig.~\ref{fig:lyapunov}, we give a Lyapnov exponent $\lambda=\lambda(c)$ defined by 
\begin{align}
\lambda=\lim_{n\to\infty} \frac1{n} \sum_{j=0}^{n-1} \log \Big| \frac{\partial }{\partial x} f(c,u_j) \Big| .
\end{align}
We can observe that $\lambda$ takes negative value for $c<0.942\cdots$ and then takes positive value.
\begin{figure}
\includegraphics[scale=0.5]{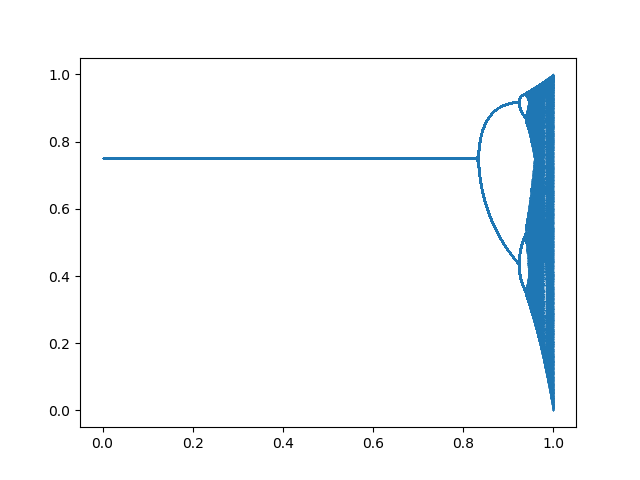}
\includegraphics[scale=0.5]{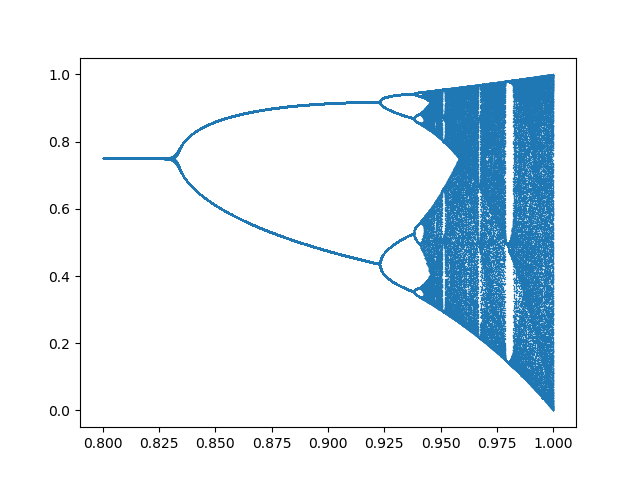}
\caption{Bifurcation diagram of the mapping~\eqref{eq:a}~~$(0.8\leq c \leq 1)$}
\label{fig:bifurcation01}
\end{figure}
\begin{figure}[htp]
\includegraphics[scale=0.5]{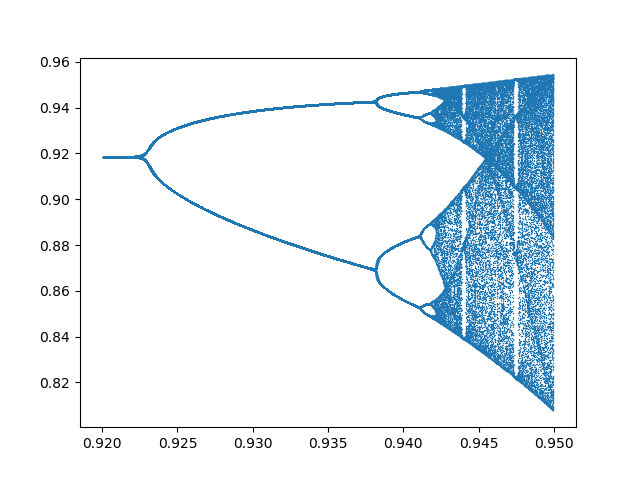}
\includegraphics[scale=0.5]{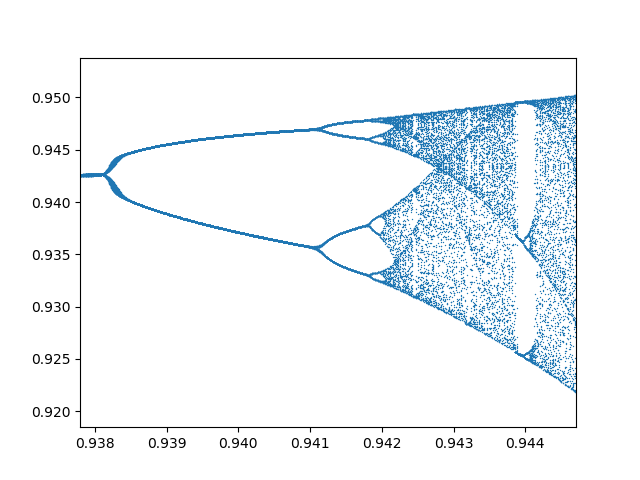}
\caption{Enlarged bifurcation diagram. (Left~:~$0.92\leq c\leq 0.95$, Right~:~$0.938\leq c \leq 0.945$)}
\label{fig:bifurcation}
\end{figure}

The corresponding bifurcation diagrams, where $(c, u_n)~~(n=200,\cdots,400)$ are 
plotted, are given in Fig.~\ref{fig:bifurcation01}. If we expand a certain region of the diagram, 
we can find a self-similar diagram, as is observed in Fig.~\ref{fig:bifurcation}. 

\section{The mapping with two parameters and 3D bifurcation diagram}
Next we extend Eq.~\eqref{eq:a} to a mapping with two parameters $A \in [3,4]$ 
and $c \in [0,1]$, given by
\begin{align}
& u_{n+1}=Au_n(1-cu_n-(1-c)u_{n+1}) \nonumber \\
\Leftrightarrow ~& u_{n+1}=\frac{Au_n(1-cu_n)}{1+A(1-c)u_n}=f(A,c,u_n),
\label{eq:twopara}
\end{align}
where $f(A,c,x)=\dfrac{Ax(1-cx)}{1+A(1-c)x}$. 

Changing the two parameters $A$ and $c $, we obtain the corresponding bifurcation 
diagrams shown in Fig.~\ref{fig:3d}, in which 
the upper and lower figures stand for the same diagram looked from different viewpoints.  
It should be noted that if $A\leqq 3$ $u_{n}$ converges to the value 
$1-\frac1{A}$ so we omitted. 

We next calculate $F_{n}=F_{n}(A)$ of Eq.~\eqref{eq:Fn}, which is given as Table \ref{t2}:

\begin{figure}[htp]
\includegraphics[scale=0.5]{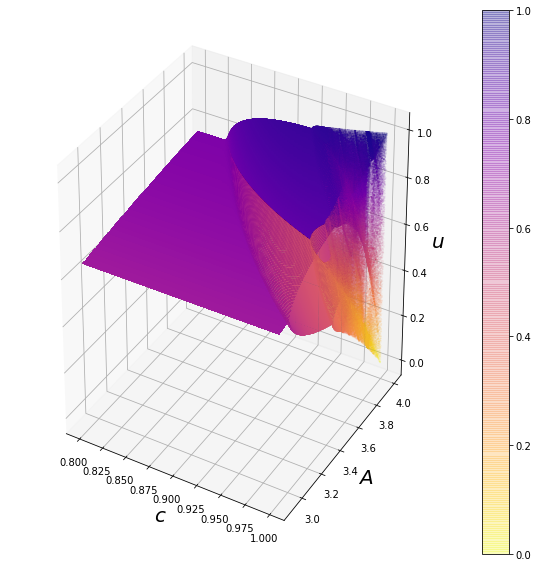}\\
\includegraphics[scale=0.5]{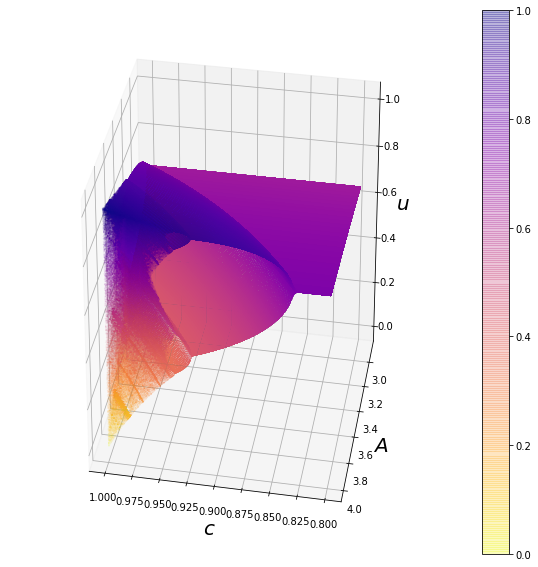}
\caption{3D bifurcation diagram corresponding to Eq.~\eqref{eq:twopara} 
from different viewpoints
}
\label{fig:3d}
\end{figure}

\begin{table}
\caption{$F_n$ in Eq.~\eqref{eq:twopara}}
\label{t2}
\begin{center}
\begin{tabular}{lllll}
\hline
\multicolumn{1}{c}{} & \multicolumn{1}{c}{$n$=1 } & \multicolumn{1}{c}{ $n$=2 } & \multicolumn{1}{c}{ $n$=3 } & \multicolumn{1}{c}{ $n$=4 } \\
\hline
$A$=3.80 & 5.89888 & 4.82824 & 4.70431 & 4.67570 \\
$A$=3.85 & 5.90474 & 4.82648 & 4.70576 & 4.67548\\
$A$=3.90 & 5.90988 & 4.82597 & 4.70410 &  4.67736 \\
$A$=3.95 & 5.91525 & 4.82493 & 4.70450 & 4.67634\\
\hline
\end{tabular}
\end{center}
\end{table}

\section{Concluding Remarks}
We have investigated a mapping with two parameters connecting the logistic mapping and the Morishita mapping, 
which are famous chaotic and integrable difference equations, respectively. It is very interesting to note 
that a peridic orbit of period 3 appears and a self-similarity of the bifurcation diagram is observed. 
These phenomena are also observed in the logistic mapping 
$u_{n+1} = A u_n(1-u_n) \quad (0 < A \leq 4)$. 
From the obtained results, we may conclude that the difference equation \eqref{eq:twopara} is a 
chaotic mapping.

\section*{Acknowledgment}
One of the authors (A. N.) is supported by JSPS KAKENHI Grant Number 18K03347.

\end{document}